\documentstyle[12pt,epsfig]{article}

\begin{document}

\begin{titlepage}

\rightline{HD-THEP-03-07}

\vskip 0.4in

\centerline{\Large\bf Kinetic approach to electroweak baryogenesis}

\vskip 1.0in

\centerline{\Large  Tomislav Prokopec$^\heartsuit$}

\vskip 0.4in

\centerline{\Large Kimmo Kainulainen$^\spadesuit$, 
        Michael G. Schmidt$^\heartsuit$ and Steffen Weinstock$^\heartsuit$}

\vskip 0.3in

\centerline{$^\heartsuit$\it
         Institut f\"ur Theoretische Physik, Universit\"at Heidelberg}

\vskip 0.1in

\centerline{\it Philosophenweg 16, D-69120 Heidelberg, Deutschland/Germany}

\vskip 0.1in

\centerline{ \it
       E-mail: T.Prokopec@, M.G.Schmidt@, S.Weinstock@thphys.uni-heidelberg.de}

\vskip 0.2in

\centerline{$^\spadesuit$ \it
         University of Jyv\"askyl\"a, PB 35 (YFL), FIN-40014, Finland}

\vskip 0.1in

\centerline{\it E-mail: kainulai@phys.jyu.fi, kainulai@nordita.dk}

\vskip .8in

\centerline{\bf Abstract}

\vskip 0.1in

After a short review of baryogenesis mechanisms, we focus on 
the charge transport mechanism at the electroweak scale, effective at
strong electroweak phase transitions. Starting from the 
one-loop Schwinger-Dyson equations for fermions coupled to 
bosons, we present a derivation of the relevant kinetic equations
in the on-shell and gradient approximations, relevant for the thick wall
baryogenesis regime. We then discuss the CP-violating source from 
the semiclassical force in the flow term, 
and compare it with the source arising in the collision term
of the kinetic equation. Finally, we summarize the results concerning the
chargino mediated baryogenesis in the Minimal Supersymmetric Standard Model.

\vskip 1.in

\footnoterule
\vskip 3truemm

{\small \noindent \uppercase{P}resented by T. Prokopec 
             at the \uppercase{I}nternational 
             \uppercase{W}orkshop
            "\uppercase{S}trong and \uppercase{E}lectroweak \uppercase{M}atter 
             2002", \uppercase{O}ctober 2-5, 2002
             \uppercase{H}eidelberg, \uppercase{G}ermany.}

\end{titlepage}


\section{Introduction}
\label{Introduction}

The necessary requirements on dynamical baryogenesis at an epoch of the early
Universe are provided by the Sakharov conditions:
(I) baryon number (B) violation,
(II) {\it charge} (C) and {\it charge-parity} (CP) violation
and (III) departure from thermal and kinetic equilibrium.
%
The Sakharov conditions may be realised at the electroweak 
transition~\cite{KuzminRubakovShaposhnikov:1985}, 
provided the transition is strongly first order. 
C and CP violation are realised in the standard model (SM) for example 
through the Cabibbo-Kobayashi-Maskawa (CKM) matrix of quarks. 
At high temperatures B is violated through sphaleron transitions,
and may be responsible for the observed baryon asymmetry today,
which is usually expressed as the baryon-to-entropy 
ratio~\cite{BurlesNollettTurner:2000+ThuanIzotov:2001},
%
  $ {n_B}/{s} \simeq {n_B}/{7n_\gamma} = 7.0 \pm 1.5 \times 10^{-11}$.
%
This is consistent both, with the nucleosynthesis, as determined
by the observed D/H ratio, and with recent cosmic microwave background 
observations~\cite{Spergel:wmap2003+Netterfield-etal:2001}.

 Over the years a large number of scenarios have been proposed to explain 
the observed matter-antimatter asymmetry of the Universe. Broadly speaking,
they can be divided into two classes: the models based on the grand 
unfication of forces, and the models based on the electroweak symmetry
breaking. In grand unified 
models~\cite{Yoshimura:1978,IgnatevKrasnikovKuzminTavkhelidze:1978}
one uses the B-L violation by the gauge 
and/or Higgs sectors of the GUT model. The particular realisations
include baryogenesis at {\it preheating} through the inflaton decay into
heavy GUT particles~\cite{KolbLindeRiotto:1996,GreeneProkopecRoos:1997}, 
whose out-of-equilibrium decays can produce a nonzero B-L; 
leptogenesis~\cite{FukugitaYanagida:1986,BuchmullerPlumacher:1996},
which involves the physics of heavy majorana neutrinos;
the Affleck-Dine mechanism~\cite{AffleckDine:1984}, which involves
the dynamics of (B-L)-violating flat directions of
supersymmetric models~\cite{DineRandallThomas:1995} (which may, but need not,
be embedded in a grand-unified model),
remnants of which may survive as B-balls~\cite{EnqvistMcDonald:1997}.

Concerning the electroweak scale baryogenesis, the Standard Model (SM) 
cannot alone be responsible for the observed matter-antimatter 
asymmetry, primarily because the LEP bound on the Higgs 
mass $m_H\ge 110~{\rm GeV}$~\cite{Heister+aleph:2002}
is inconsistent with the requirement that the 
transition be strongly first order, in order for the baryons
produced in the symmetric phase be not washed-out 
by the sphaleron transitions in the Higgs (`broken') phase.

 Supersymmetric extensions of the Standard Model, on the other hand,
may result in a strong first order transition. Indeed, in the Minimal
Supersymmetric Standard Model the sphaleron bound can be satisfied,
provided the stop and the lightest Higgs particles are not too heavy, 
$m_{\tilde t}\leq 170$~GeV and $m_{H}\le 120$~GeV~\cite{Quiros:2001}.
Non-minimal supersymmetric extensions can typically provide stronger
phase transitions, and are hence less constrained~\cite{HuberSchmidt:2000}.

 An efficient mechanism for baryon production at the electroweak phase 
transition is the 
{\it charge transport mechanism}~\cite{CohenKaplanNelson:1991}, which works
as follows: At a first order transition, when the Universe supercools,
the bubbles of the Higgs phase nucleate and grow. In presence of
a CP-violating condensate at the bubble interface, 
as a consequence of collisions of chiral fermions with 
scalar particles in presence of a scalar field condensate,
CP-violating currents are created
and transported into the symmetric phase, where they bias baryon number
production. The baryons thus produced are transported back into the Higgs
phase where they are frozen-in. The main unsolved problem of electroweak 
baryogenesis is the systematic computation of the relevant CP-violating
currents generated at the bubble interface. Here we shall reformulate this 
problem in terms of calculating CP-violating sources in the kinetic
Boltzmann equations for fermions.

The techniques we report here are relevant for calculation of 
sources in the limit of thick phase boundaries and a weak coupling
to the Higgs condensate, which are both generically realised 
in supersymmetric models. In this case one can show that, to linear order
in the Planck constant $\hbar$, the quasiparticle picture for fermions
survives~\cite{KainulainenProkopecSchmidtWeinstock-I,KainulainenProkopecSchmidtWeinstock-II}. In presence of a CP-violating 
condensate there are two types of sources: the semiclassical force 
in the flow term of the kinetic Boltzmann equation, and the collisional 
sources. The semiclassical force was originally introduced for baryogenesis
in two-Higgs doublet models~\cite{JoyceProkopecTurok:1994},
and subsequently adapted to the chargino baryogenesis in the
Minimal Supersymmetric Standard Model 
(MSSM)~\cite{ClineJoyceKainulainen:2000}. The semiclassical force corresponds
to tree-level interactions with a semiclassical background field,
and it is universal in that its form is independent on interactions. 
The collisional sources, on the other hand, arise when fermions in the loop
diagrams interact with scalar background fields. These sources arise 
first from the one-loop diagrams, in which fermions interact with a 
CP-violating scalar background. When viewed in the kinetic Boltzmann equation, 
these processes correspond to tree-level interactions in which fermions
absorb or emit scalar particles, whilst interacting in a CP-violating 
manner with the scalar background. The precise form of the collisional
sources depends on the form of the interaction. In the following sections
we discuss how one can study the CP-violating collisional sources induced
by a typical Yukawa interaction term.

\section{Kinetic equations and the quasiparticle picture}

Here we work in the simple model of chiral fermions coupled to
a complex scalar field {\it via} the Yukawa 
interaction~\cite{KainulainenProkopecSchmidtWeinstock-I,KainulainenProkopecSchmidtWeinstock-II}  
%
%
%
\begin{equation}
{\cal L}_{\rm Yu} = - y\phi \bar{\psi}_L\psi_R
           - y \phi^* \bar{\psi}_R\psi_L,
\label{lagrangian1}
\end{equation}
which, at a phase transition, may give rise to a complex, 
spatially varying, mass term
\begin{equation}
     m(u) \equiv y' \Phi_0(u) = m_R(u) + i m_I(u) 
                              = |m(u)|\mbox{e}^{i\theta(u)}
.
\label{mass1}
\end{equation}
Here $\Phi_0(u) = \langle \Omega| \hat \Phi(u)|\Omega\rangle$, and
$|\Omega\rangle$ is the physical state.
Such a situation is realised, for example, by the Higgs field at a first
order electroweak phase transition.

\begin{figure}[ht]
\centerline{\epsfxsize=4.9in\epsfbox{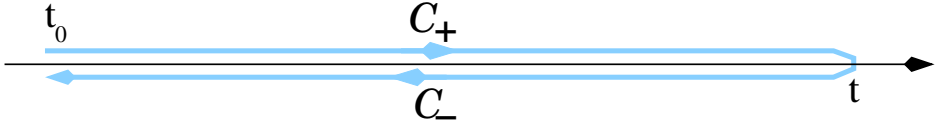}}   
\vskip -0.1in
\caption{\small The closed time contour for the Schwinger-Keldysh
         nonequilibrium formalism.
         \label{figure-I}
        }
\end{figure}

The dynamics of quantum fields can be studied by considering the equations 
of motion arising from the two-particle irreducible (2PI) effective action 
\cite{LuttingerWard:1960+Baym:1962,CornwallJackiwTomboulis:1974} in the Schwinger-Keldysh
closed-time-path formalism~\cite{Schwinger:1961,ChouSuHaoYu:1985}. 
This formalism is suitable for studying the dynamics of the non-equilibrium
fermionic and bosonic two-point functions 
\begin{eqnarray}
iS_{\alpha\beta}(u,v) 
 &=& \langle\Omega|T_{{\cal C}}
         \big[\psi_\alpha(u)\bar{\psi}_\beta(v)\big]|\Omega\rangle
\label{S}
\\
i\Delta(u,v) 
 &=& \langle\Omega|T_{{\cal C}}\big[\phi(u){\phi}^\dagger(v)\big]|\Omega\rangle
,
\label{Delta}
\end{eqnarray}
where the time ordering
$T_{{\cal C}}$ is along the Schwinger contour shown in 
figure~\ref{figure-I}. 
The ${\cal C}$-time ordering 
can be conveniently represented in the Keldysh component formalism.
For example, for nonequilibrium dynamics of quantum fields 
the following Wightman propagators are of particular relevance,  
\begin{eqnarray}
iS^<(u,v) &=& - \langle\Omega|\bar{\psi}(v)\psi(u)|\Omega\rangle
\nonumber\\
i\Delta^<(u,v) &=& \;\;\, \langle\Omega|\phi^\dagger(v)\phi(u)|\Omega\rangle.
\label{Wightman}
\end{eqnarray}

For thick walls, {\it i.e.} when the de Broglie wavelength $\ell_{\rm dB}$
of typical plasma excitations is small in comparison to the phase interface
thickness $L_w$, it is suitable to work in the Wigner representation for
the propagators, which corresponds to the Fourier transform with respect
to the relative coordinate $r=u-v$, and expand in the gradients of
average coordinate $x=(u+v)/2$. This then represents an expansion in powers
of $\ell_{\rm dB}/L_w$. When written in this Wigner representation, 
the kinetic equations for fermions
become~\cite{KainulainenProkopecSchmidtWeinstock:2002c}
\begin{equation}
{\cal D}S^< \equiv 
   \Big(\frac i2\partial\!\!\!/+ k\!\!\! / 
 -(mP_R-m^*P_L) e^{{-\frac i2\stackrel{\leftarrow}{\partial}}\cdot\;\partial_k}
   \Big) S^{<}
    = {\cal C}_\psi,
\label{eom-S<}
\end{equation}
where for simplicity we neglected the contributions from self-energy 
corrections to the mass and the collisional broadening term,
which is of the form $-{\rm e}^{-i\diamond}\{\Sigma^{<}\}\{S_h\}$,
where $S_h = (S^r+S^a)/2$ is the hermitean part of the propagator, 
and $S^r$ and $S^a$ denote the retarded and advanced propagators, 
respectively. 
By considering, as an example, the scalar field theory, 
we have been able to show that, to first order in gradients, 
the flow term can be rewritten 
as~\cite{KainulainenProkopecSchmidtWeinstock:2003} 
\begin{equation}
         {\cal A}_s\; \diamond \{\Omega_\phi^2\}\{n_\phi\}
     - 2\Gamma_\phi{\cal A}_s\Delta_h \;
           \diamond \{\Gamma_\phi\}\{n_\phi\}  
     = {\cal C}_\phi
,
\label{collisional broadening}
\end{equation}
where ${\cal A}_s\propto \delta (k^2 - m_\phi^2-\Sigma_h)$ 
denotes the on-shell spectral function,
$n_\phi$ the bosonic occupation number,  
$\Delta_h = (\Delta^r + \Delta^a)/2$
and ${\cal C}_\phi$ the scalar collision term. 
This implies that, when working to {\it linear} order in
the width $\Gamma_\phi = (i/2)(\Pi^>-\Pi^<)$ 
and self-energy $\Pi_h = (\Pi^r+\Pi^a)/2$, 
one can include the self-energy and collision term, while the effect
of the collisional broadening, described by the second
term in~(\ref{collisional broadening}), can be consistently neglected.
While we focus here on the effects of 
a pseudoscalar mass, one should keep in mind that
one can include the effect of the self-energy within the on-shell 
approximation, provided one appropriately modifies the spectral condition.

\begin{figure}[ht]
\centerline{\epsfxsize=4.9in\epsfbox{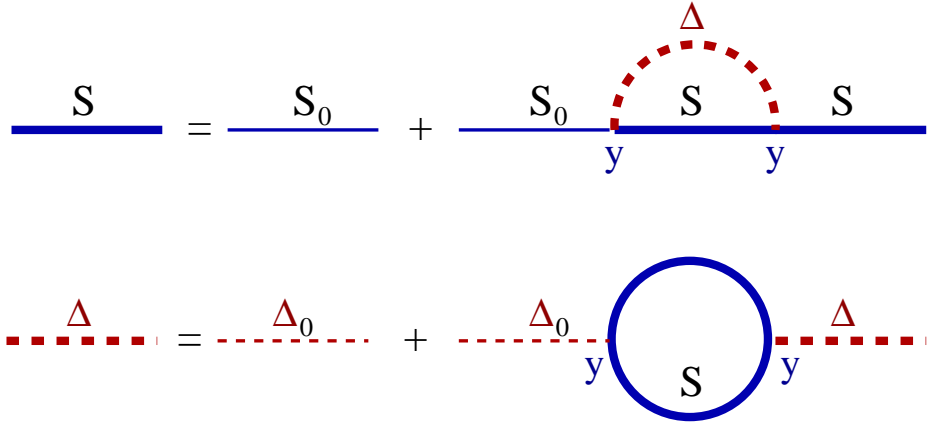}}   
\vskip -0.1in
\caption{\small The one-loop Schwinger-Dyson equations for the 
         out-of-equilibrium fermionic ($S$) and scalar ($\Delta$) propagators. 
         When projected on-shell and expanded in gradients, 
         these equations reduce to the kinetic Boltzmann equations.
        \label{figure-II}
        }
\end{figure}

When the collision term ${\cal C}_\psi$ is approximated at the one-loop
(with the resummed propagators),
equation~(\ref{eom-S<}) corresponds to the nonequilibrium 
fermionic Schwinger-Dyson equation shown in figure~\ref{figure-II}. Since
the flow term of the scalar equation (also shown in figure~\ref{figure-II})
does not yield CP-violating sources at first order in
gradients~\cite{KainulainenProkopecSchmidtWeinstock-I,KainulainenProkopecSchmidtWeinstock:2003}, we do not discuss it here.

As the bubbles grow large, they tend to become more and more planar. Hence,
it suffices to consider the limit of a planar phase interface, in which 
the mass condensate in the wall frame becomes a function of
one coordinate only, $m=m(z)$. Further, we keep only 
the terms that contribute at order $\hbar$ to Eq.~(\ref{eom-S<}),
which implies that we need to keep second order gradients of the mass term
\begin{eqnarray}
m e^{{-\frac i2\stackrel{\leftarrow}{\partial}}\cdot\;\partial_k}
   = m + \frac i2 m' \partial_{k_z}  
    -  \frac 18 m'' \partial_{k_z}^2 + o(\partial_z^3), 
\label{m-hat}
\end{eqnarray}
where $m=m(z)$, $m' \equiv \partial_z m$ and $m'' \equiv \partial_z^2 m$. 
On the other hand, in the collision term  ${\cal C}_\psi$ we need to 
consider terms only up to linear order in derivatives 
\begin{eqnarray}
 {\cal C}_\psi  &=& {\cal C}_{\psi 0} + {\cal C}_{\psi 1} + ..
 \nonumber\\ 
  {\cal C}_{\psi 0} &=& - \frac 12
           \Big(\Sigma^>S^< - \Sigma^<S^> \Big)
 \nonumber\\ 
  {\cal C}_{\psi 1} &=& - \frac i4 
 (\partial_z^{(1)}\partial_{k_z}^{(2)} - \partial_{k_z}^{(1)}\partial_z^{(2)})
             \Big(\Sigma^>S^< - \Sigma^< S^>\Big),
\label{C-psi}
\end{eqnarray}
where $\Sigma^<$ and $\Sigma^>$ represent the fermionic self-energies, and 
the derivatives $\partial_{z}^{(1)}$, $\partial_{k_z}^{(1)}$ 
($\partial_z^{(2)}$, $\partial_{k_z}^{(2)}$) act on the first
(second) factor in the parentheses. 

 An important observation is that,
when $G = G(k_\mu,t-\vec x_\|\cdot\vec k_\|,z)$, the spin 
\begin{equation}
  S_z  \equiv L^{-1}(\Lambda) \,\tilde{\!S}_z L(\Lambda)
  \; = \;  
 \gamma_\| \Big( \tilde S_z 
      - i(\vec v_\|\times \vec \alpha)_z \Big) 
\label{Sz}
\end{equation}
is conserved 
\begin{equation}
 [{\cal D},S_z]S^< = 0,
\label{spin-commutator}
\end{equation}
where ${\cal D}$ is the differential operator in Eq.~(\ref{eom-S<}),
$\vec\alpha = \gamma^0 \vec\gamma$, $\tilde S_z = \gamma^0\gamma^3\gamma^5$,
and $\gamma_\| = 1/(1- \vec v_\|^{\,2})^{1/2}$. The spin operator correponds to 
the boosted spin in the $z$-direction (of the interface motion), and 
when written as the Pauli-Lubanski spin operator,
\begin{equation}
  S_{PL}(k,s) \equiv - \frac{1}{e_0}{\mathbin{k\mkern-9.8mu\big/}}
                                    {\mathbin{s\mkern-9mu\big/}}\gamma^5, 
\qquad e_0 \equiv (k^2)^{1/2},
\quad s^2 = -1,
\quad s\cdot k = 0
,
\label{Pauli-Lubanski}
\end{equation}
the spin 4-vector corresponds to 
\begin{equation}
  s^\mu = \frac{1}{\tilde k_0 e_0}\left( 
        \begin{array}{c}  k_0 k_z \\   
                          k_x k_z \\   
                          k_y k_z \\   
                          \tilde k_0^2
        \end{array} 
      \right).
\label{spin-direction}
\end{equation}
In the highly relativistic limit we have $\tilde k_0^2 \approx k_z^2$, 
and the spin vector $\vec s \propto \vec k$, such that 
the spin operator $S_z$ approaches the helicity operator,
\begin{eqnarray}
  \hat H (\vec k) &=& -\frac{1}{e_0}{\mathbin{k\mkern-9.8mu\big/}}
                                     {\mathbin{h\mkern-10mu\big/}}\gamma^5 
\nonumber\\
         &=& \hat{\vec k}\cdot \gamma^0 \vec\gamma\gamma^5,
\qquad  
       h^\mu = \frac{1}{e_0}\left( 
        \begin{array}{c}  |\vec k| \\   
                           k_0 \hat{\vec k}
        \end{array} 
      \right),
\label{helicity-operator}
\end{eqnarray}
as one would expect. (As usually, the helicity operator measures spin in the 
direction of particle's motion, $\hat{\vec k} = \vec k/|\vec k|$.)
As a consequence, 
for light particles with momenta of order the temperature, 
$k\sim T \gg m$, the spin states we consider here can be approximated
by the helicity states, which are often used in literature for baryogenesis
calculations~\cite{JoyceProkopecTurok:1994,ClineJoyceKainulainen:2000}. 
To answer the question to what extent is this fulfilled,
requires a detailed quantitative study, which is beyond the scope of 
this talk~\cite{KainulainenProkopecSchmidtWeinstock:2003}.

This discussion implies that, without loss of generality, 
the fermionic Wigner function can be written in the following
block-diagonal form 
\begin{eqnarray}
    S^< &=& \sum_{s=\pm} S_s^<
\nonumber\\
    S^< &=& L(\Lambda)^{-1}\tilde S^< L(\Lambda)
\nonumber\\
    - i\gamma^0\tilde{S}^{<}_s 
    &=& \frac{1}{4} 
   ({\mathbf{1}}+s\sigma^3) \otimes \rho^a\tilde{g}^s_a,
\label{Dec1+1}
\end{eqnarray}
where $\sigma^3$ and $\rho^i$ ($i=1,2,3$) are the Pauli matrices,
$\rho^0 = {\mathbf{1}}$ is the $2\times 2$ unity matrix,
and $L(\Lambda)$ is the following Lorentz boost operator
\begin{equation}
  L(\Lambda) = \frac{k_0 + \tilde{k}_0 
               - \gamma^0\vec{\gamma}\cdot\vec{k}_{\|}}
                {\sqrt{2\tilde{k}_0(k_0+\tilde{k}_0)}},
\label{L}
\end{equation}
with $\tilde k_0 = {\rm sign}(k_0)({k_0^2 -\vec k_\|^2})^{1/2}$, and 
$\Lambda$ corresponds to the Lorentz boost that transforms away $\vec k_\|$. 

 With the decomposition~(\ref{Dec1+1}) the trace of the antihermitean part
of Eq.~(\ref{eom-S<}) can be written as the following 
{\em algebraic} constraint 
equation~\cite{KainulainenProkopecSchmidtWeinstock-II} 
\begin{equation}
\Bigl(k^2 - |m|^2 + \frac{s}{\tilde{k}_0}\,|m|^2\theta'
    \Bigr) g^s_{00} = 0 ,
\label{constraint3+1}
\end{equation}
where $g^s_{00} = \gamma_\| \tilde g_0^{s}$ denotes
the particle density on phase space $\{k_\mu,x_\nu\}$.
Equation~(\ref{constraint3+1}) has a spectral solution

\begin{eqnarray}
g^{s}_{00}  &\equiv& \sum_\pm \frac{2\pi }{Z_{s\pm}}
                 \, n_s \,\delta(k_0 \mp \omega_{s\pm}),
\label{spectral-dec}
\end{eqnarray}
where $\omega_{s\pm}$ denotes the dispersion relation
\begin{equation}
     \omega_{s\pm} = \omega_0 \mp s \frac{|m|^2\theta'}
   {2\omega_0\tilde\omega_0},
     \qquad \qquad 
\omega_0 = \sqrt{ \vec k^2 + |m|^2},
\qquad
\tilde\omega_0 = \sqrt{\omega_0^2-\vec k_{\parallel}^{\,2}}
\label{dispersion1}
\end{equation}
and $Z_{s\pm} = 
  1 \mp s|m|^2\theta'/2\tilde\omega_0^{3}$. The delta functions 
in~(\ref{spectral-dec}) project $n_s(k_\mu,t-\vec x_\|\cdot\vec k_\|,z)$ 
on-shell, thus yielding the distribution functions $f_{s+}$ and $f_{s-}$
for particles and antiparticles with spin $s$, respectively, defined by 
\begin{eqnarray}
     f_{s+} &\equiv& n_s(\omega_{s+},k_z,t-\vec x_\|\cdot\vec k_\|,z) \nonumber
 \\
     f_{s-} &\equiv& 1 - n_s(-\omega_{s-},-k_z,t+\vec x_\|\cdot\vec k_\|,z).
\label{fs}
\end{eqnarray}
This on-shell projection proves the implicit
assumption underlying the semiclassical WKB-methods, that the plasma
can be described as a collection of single-particle excitations with
a nontrivial space-dependent dispersion relation. 
In fact, the decomposition~(\ref{Dec1+1}), Eq.~(\ref{constraint3+1}) and 
the subsequent discussion imply that the physical states that correspond
to the quasiparticle plasma excitations are the eigenstates of the spin
operator~(\ref{Sz}). 

 Taking the trace of the Hermitean part of Eq.~(\ref{eom-S<}), integrating
over the positive and negative frequencies and taking account
of~(\ref{spectral-dec}) and~(\ref{fs}), one obtains 
the following on-shell kinetic 
equations~\cite{KainulainenProkopecSchmidtWeinstock-II}
\begin{equation}
        \partial_t f_{s\pm} + \vec v_\|\cdot \nabla_\| f_{s\pm}
      +  v_{s\pm} \partial_z f_{s\pm}
      + F_{s\pm} \partial_{k_z} f_{s\pm} = {\cal C}_{\psi s\pm}[f_{s\pm}],
\label{ke-fspm0}
\end{equation}
where $f_{s\pm} = f_{s\pm}(\vec k, z, \, t - \vec v_\| \cdot \vec x_\|)$,
${\cal C}_{s\pm}[f_{s\pm}]$ is the collision term obtained by 
integrating~(\ref{C-psi}) over the positive and negative frequencies, 
respectively, the quasiparticle {\it group velocity} 
$v_{s\pm} \equiv k_z/\omega_{s\pm}$ is expressed in terms of the kinetic
momentum $k_z$ and the quasiparticle energy 
$\omega_{s\pm}$~(\ref{dispersion1}), and the {\em semiclassical force} 
\begin{eqnarray}
    F_{s\pm} = - \frac{{|m|^2}^{\,\prime} }{2\omega_{s\pm}}
                   \pm  \frac{s(|m|^2\theta^{\,\prime})^{\,\prime}}
                   {2\omega_0\tilde\omega_0}.
\label{Fspm}
\end{eqnarray}
In the stationary limit in the wall frame the distribution function
simplifies to $f_{s\pm} = f_{s\pm}(\vec k, z)$.
When compared with the 1+1 dimensional 
case~\cite{KainulainenProkopecSchmidtWeinstock-I}, the sole, but significant,
difference in the force~(\ref{Fspm}) is that the CP-violating 
$|m|^2\theta'$-term is enhanced by the boost-factor $\gamma_\parallel 
= \omega_0/\tilde\omega_0$,
$\tilde \omega_0 = ({\omega_0^2-\vec k_\parallel^{\,2}})^{1/2}$, which,
when integrated over the momenta, leads to an enhancement by about a 
factor {\it two} in the CP-violating source from the semiclassical force. 

\section{Sources for baryogenesis in the fluid equations}

\begin{figure}[ht]
\vskip -0.1in
\centerline{\epsfxsize=4.in\epsfbox{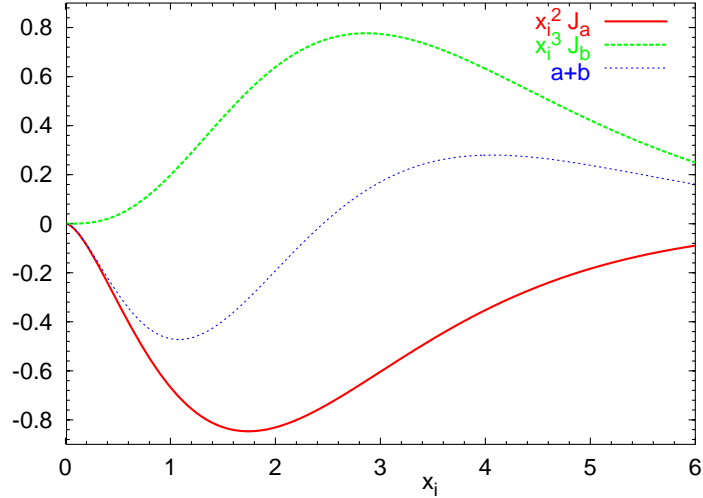}}   
\vskip -0.15in
\caption{\small The flow term 
         sources~(\ref{boltzeqn-split-6a})-(\ref{boltzeqn-split-6b}) 
         characterised by the integrals $x_i^2{\cal J}_a(x_i)$ 
         ({\it red solid}) and $x_i^3{\cal J}_b(x_i)$ ({\it green dashed})
         as a function of the rescaled mass $x_i=|m_i|/T$. 
         The sum of the two sources ({\it dotted blue})
         is also shown.
         \label{figure-III}
        }
\end{figure}
Fluid transport equations are usually obtained by taking the first two
moments of the Boltzmann transport equation~(\ref{ke-fspm0}):
integrating~(\ref{ke-fspm0}) over the spatial momenta results in 
the {\it continuity} equation for the vector current, while multiplying by 
the velocity and integrating over the momenta yields the {\it Euler} equation.
The physical content of these equations can be summarized as the particle
number and fluid momentum density conservation laws for fluids, respectively. 
This procedure is necessarily approximate simply because the fluid equations
describe only very roughly the rich momentum dependence described by the 
distribution functions of the Boltzmann equation~(\ref{Fspm}). 
The fluid equations can be easily reduced to the diffusion equation
which has so far being used almost exclusively for electroweak baryogenesis
calculations at a first order electroweak phase transition. A useful 
intermediate step in the derivation of the fluid equations is
rewriting Eq.~(\ref{ke-fspm0}) for the CP-violating departure from equilibrium
$\delta f_{si} = \delta f_{si+} - \delta f_{si-}$ as follows
\begin{eqnarray}
  &&  \left(\partial_t  + \frac{k_z}{\omega_{0i}} \partial_z 
  - \frac{{|m_i|^2}'}{2\omega_{0i}} \partial_{k_z}\right) \delta f_{si} 
  + v_w \delta F_{si} (\partial_{\omega} f_{\omega})_{\omega_{0i}} 
\nonumber\\
  &+& v_w F_{0i}\delta\omega_{si}
 \bigg[\Big(\frac{\partial_{\omega} f_{\omega}}{\omega}\Big)_{\omega_{0i}}
      - (\partial^2_{\omega} f_{\omega})_{\omega_{0i}} \bigg]
  = {\cal C}_{\psi si},
\label{boltzeqn-split-2}
\end{eqnarray}
where $i$ is the species (flavour) index, 
$f_\omega = 1/(e^{\beta\omega}+1)$, and 
%
\begin{eqnarray}
    F_{0i} &=& - \frac{{|m_i|^2}'}{2\omega_{0i}}
\nonumber\\
 \delta\omega_{si} &=& s\frac{(|m_i|^2\theta_i)'}{\omega_{0i}\tilde\omega_{0i}}
\nonumber\\
\delta F_{si} &\equiv& F_{si+} - F_{si-} 
              = s\frac{(|m_i|^2\theta_i')'}{\omega_{0i}\tilde\omega_{0i}}
\nonumber\\
   {\cal C}_{\psi si} &\equiv& {\cal C}_{\psi si+} - {\cal C}_{\psi si-}.
\label{boltzeqn-split-3}
\end{eqnarray}
\begin{figure}[ht]
\vskip -0.1in
\centerline{\epsfxsize=4.9in\epsfbox{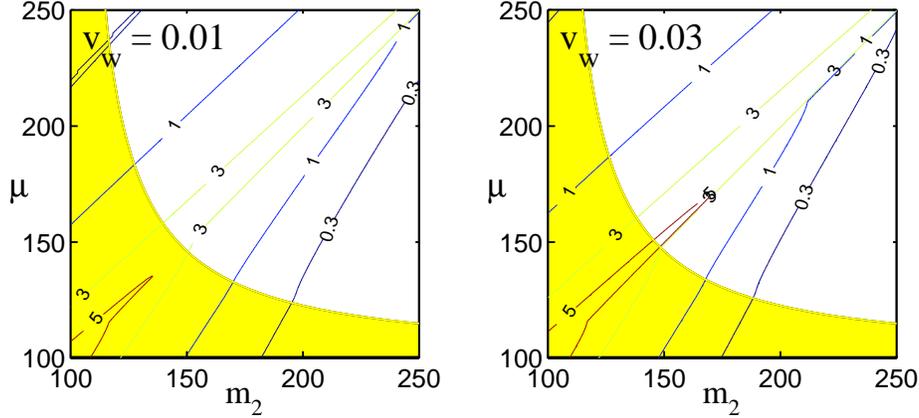}}
\vskip -0.65in
\caption{\small The semiclassical force baryogenesis mediated by charginos
         of the MSSM calculated in the helicity basis. The figure shows
         contours for the baryon-to-entropy ratio in the units of $10^{-11}$
         for two wall velocities $v_w = 0.01$ and $v_w = 0.03$ as
         a function of the soft susy breaking parameters $\mu$ and $m_2$.
         A maximal CP violation in the chargino sector is assumed.
         The shaded (yellow) regions are ruled out by the LEP measurements. 
         The observed baryon asymmetry is in these units $5-9$. 
         (The figure is taken from the latter reference 
         in~\cite{ClineJoyceKainulainen:2000}).
         \label{figure-IV}
        }
\vskip -0.in
\end{figure}
When integrating~(\ref{boltzeqn-split-2}) over the momenta, 
the flow term yields two sources in the continuity equation for the vector
current. The former comes from the CP-violating spin dependent semiclassical
force, and has the form
\begin{eqnarray}
 {\cal S}^a_{si} &=& v_w \int \frac{d^3k}{(2\pi)^3}\delta F_{si} 
             (\partial_{\omega} f_{\omega})_{\omega=\omega_{0i}}
\nonumber\\
               &=& -sv_w \frac{(|m_i|^2\theta'_i)'}{4\pi^2}\; {\cal J}_a(x_i)
,
\label{boltzeqn-split-6a}
\end{eqnarray}
with $x_i=|m_i|/T$, while the latter comes from the CP-violating shift
in the quasiparticle energy, and can be written as 
\begin{eqnarray}
 {\cal S}^b_{si} &=& v_w \int \frac{d^3k}{(2\pi)^3}
F_{0i}\delta\omega_{si}
 \Big[\big(\partial_{\omega} f_{\omega}/\omega\big)_{\omega_{0i}}
      - (\partial^2_{\omega} f_{\omega})_{\omega_{0i}} \Big]
\nonumber\\
        &=& sv_w \frac{(|m_i|^2\theta'_i)}{2\pi^2T}{|m_i|}'\;{\cal J}_b(x_i).
\label{boltzeqn-split-6b}
\end{eqnarray}
The total source is simply the sum of the two,
${\cal S}_{si} = {\cal S}^a_{si} + {\cal S}^b_{si}$.
To get a more quantitative understanding of these sources, in
figure~\ref{figure-III} we plot the integrals ${\cal J}_a$ and ${\cal J}_b$
in equations~(\ref{boltzeqn-split-6a}) and~(\ref{boltzeqn-split-6b}).
A closer inspection of the sources ${\cal S}_{si}^a$ and ${\cal S}_{si}^b$
indicates that the total source ${\cal S}_{si}$ can be also rewritten as
the sum of two sources: the source $\propto {|m_i|^2}'\theta_i'$, 
characterized by $x_i^2{\cal J}_a + x_i^3 {\cal J}_b$, and the source
$\propto {|m_i|^2}\theta_i''$, characterized by $x_i^2{\cal J}_a$.
We note that, in the spin state quasiparticle basis the flow term
sources appear in the continuity equation for the vector current,
while in the helicity basis, which is usually used in 
literature~\cite{JoyceProkopecTurok:1994,ClineJoyceKainulainen:2000}, 
the flow term sources appear in the Euler equation. 
In figure~\ref{figure-IV} we show recent results of baryogenesis 
calculations of Ref.~\cite{ClineJoyceKainulainen:2000} based on the 
CP-violating contribution to the semiclassical force in the chargino sector
of the Minimal Supersymmetric Standard Model (MSSM). This calculation is 
based on the quasiparticle helicity states picture, and 
it suggests that one can dynamically obtain
baryon production marginally consistent with the observed
value, ${n_B}/{s} = 7.0 \pm 1.5 \times 10^{-11}$, 
provided $m_2\sim \mu \sim 150$~GeV and $v_w \sim 0.03$ ($c = 1$).

\begin{figure}[ht]
\vskip -0.in
\centerline{\hskip 0.38in\epsfxsize=3.6in\epsfysize=3.3in\epsfbox{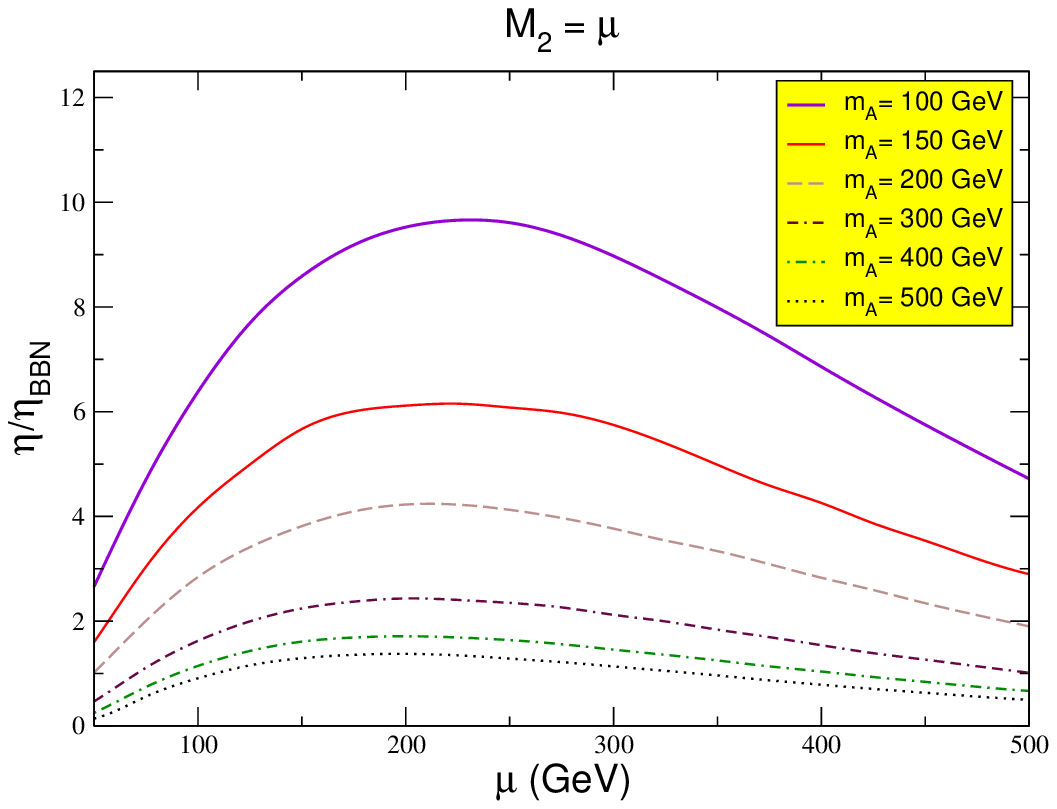}
            \hskip -0.33in\epsfxsize=3.6in\epsfysize=3.3in\epsfbox{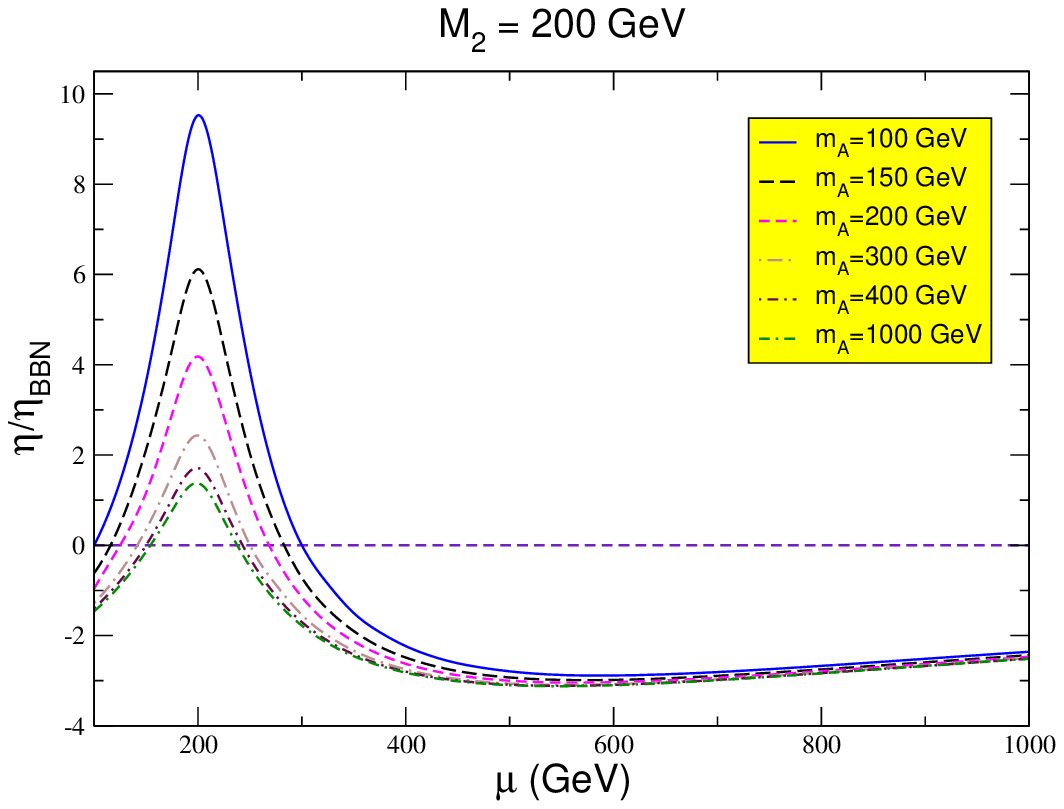}}
       \vskip -0.2in
\caption{\small Baryon production mediated by charginos of the MSSM 
         from Ref.~\cite{CarenaQuirosSecoWagner:2002},
         expressed as the multiple of the observed value, 
         $\eta_{BBN} = n_B/s \simeq 6\times 10^{-11}$.
         The figures show contours for the baryon-to-photon ratio
         in the units of $10^{-10}$ as a function of $\mu$ and $m_2$
         for $\tan \beta =10$, (a) $m_2 = \mu$ and (b) $m_2 = 200~{\rm GeV}$.
         The wall velocity is taken to be $v_w = 0.01$, and 
         the maximum CP violation in the chargino sector 
         is assumed.
         \label{figure-V}
        }
\end{figure}

At this moment, the question of baryogenesis mediated {\it via} the charginos 
of the MSSM is not completely resolved. Indeed, more 
recenly the results, which we summarise in figure~\ref{figure-V},
have been reported~\cite{CarenaQuirosSecoWagner:2002},
where the relevant CP-violating sources were computed in the flavour basis.
Flavour mixing 
was, however, not taken account of, which is, on the wall, formally 
of the order $\hbar^0$, and hence cannot be neglected. 

 In conclusion we note that, even though we have recently witnessed 
important progress modelling dynamical baryon production in 
supersymmetric models, some important questions remain unresolved
which may have quantitative impact on the final results for baryon production.

\begin{figure}[ht]
\vskip -0.0in
\centerline{\epsfxsize=4.4in\epsfbox{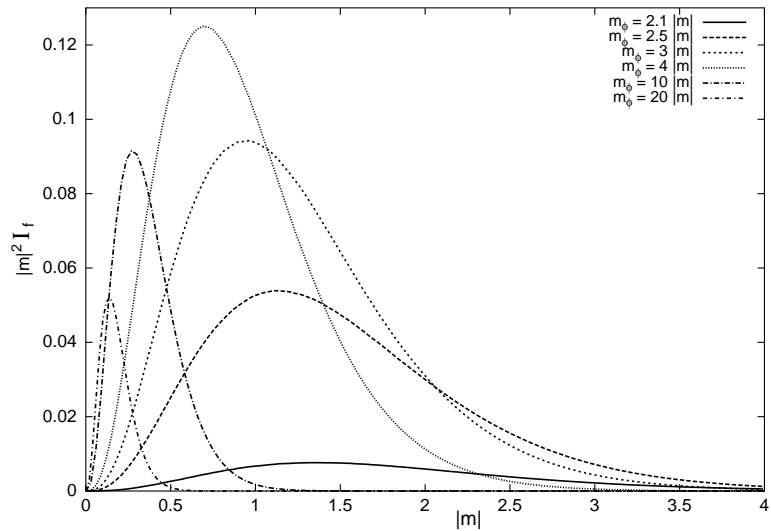}}   
\vskip -0.45in
\caption{\small The collisional source contributing to the fermionic 
         kinetic equation at one loop for the mass ratios
         $m_\phi/|m| = 2.1, 2.5, 3, 4, 10$ and 20, respectively. 
         The source peaks for $|m| \approx 0.7T$ and $m_\phi\approx 4|m|$.
         \label{figure-VI}
        }
\vskip -0.in
\end{figure}

 We now turn to discussion of the collision term sources 
in Eqs.~(\ref{ke-fspm0}) and~(\ref{boltzeqn-split-2}). 
We assume that the self-energies $\Sigma^{>,<}$ are approximated 
by the one-loop expressions ({\it cf.} figure~\ref{figure-II})
\begin{eqnarray}
&& \Sigma^{<,>}(k,x) =
\nonumber\\
&& i y^2\int \frac{d^4k'd^4k''}{(2\pi)^8}\,\big[
(2\pi)^4\delta(k-k'+k'') P_L S^{<,>}(k',x) P_R \Delta^{>,<}(k'',x) 
\nonumber\\
&& \qquad\qquad\qquad
  + (2\pi)^4\delta(k-k'-k'') P_R S^{<,>}(k',x) P_L \Delta^{<,>}(k'',x)
\big],\qquad
\label{Sigma-<>}
\end{eqnarray}
where $\Delta^{<}$ and $\Delta^{>}$ denote the bosonic Wigner functions.
This expression contains both the CP-violating sources and 
relaxation towards equilibrium. The CP-violating sources can be 
evaluated by approximating the Wigner functions 
$S^{>,<}$ and $\Delta^{>,<}$ by the equilibrium expressions accurate to 
first order in derivatives. The results of the investigation are as follows.
There is no source contributing to the continuity equation, while the source
arising in the Euler equation is of the 
form~\cite{KainulainenProkopecSchmidtWeinstock:2002c}
\begin{eqnarray}
2\int_{\pm}\frac{d^4k}{(2\pi)^4}\frac{k_z}{\omega_0} {\cal C}_{\psi si} 
 &=& v_w y^2\frac { s|m|^2\theta'}{32\pi^3 T}
{\cal I}_f(|m|,m_\phi),
\label{FermionicSource1}
\end{eqnarray}
where the function ${\cal I}_f(|m|,m_\phi)$ is plotted in 
figure~\ref{figure-VI}. It is encouraging that the source vanishes 
for small values of the mass parameters, which suggests that the expansion
in gradients we used here yields the dominant sources. Note that the 
source is nonvanishing only in the kinematically allowed region, 
$m_\phi\geq 2 |m|$. When the masses are large, $|m|, m_\phi \gg T$, 
the source is, as expected, Boltzmann-suppressed. 
It would be of interest to make a comparison between the sources
in the flow term and those in the collision 
term, and apply our methods to realistic 
models~\cite{KainulainenProkopecSchmidtWeinstock:2003}.

\vskip 0.2in

%
%

\end{document}